\documentclass[twocolumn,showpacs,floatfix,prb,10,aps]{revtex4-1}
\usepackage{amsmath,amssymb}
\usepackage{epsfig,psfrag}
\usepackage{dcolumn}
\usepackage{bm}
\usepackage{graphicx}
\usepackage{color}

\begin{document}

\title{
Quasienergy description of the driven  Jaynes-Cummings model}
\author{V. Peano$^1$ and M. Thorwart$^{2}$}
\affiliation{
$^1$Freiburg Institute for Advanced Studies (FRIAS),
Albert-Ludwigs-Universit\"at Freiburg,
79104 Freiburg, Germany \\
$^2$I.\ Institut f\"ur Theoretische Physik,  Universit\"at Hamburg,
Jungiusstra{\ss}e 9, 20355 Hamburg, Germany
} 
\date{\today}

\begin{abstract}
We analyze the driven resonantly coupled Jaynes-Cummings model in terms of a
quasienergy approach by switching  to a frame rotating with the external
modulation frequency and by using the dressed atom picture. A quasienergy
surface in phase space emerges whose level spacing is governed by a rescaled
effective Planck constant. Moreover, the well-known multiphoton transitions can
be reinterpreted as resonant tunneling transitions
from the local maximum of the quasienergy surface.
Most importantly, the driving defines a quasienergy well  which is
nonperturbative in nature. The quantum mechanical quasienergy state localized at
its bottom is squeezed. 
In the Purcell limited regime, the potential well is metastable and the
effective local
temperature close to its minimum is uniquely determined by
the squeezing factor. The activation occurs in this case via dressed spin flip
transitions rather
than via quantum activation as in other driven nonlinear quantum systems such
 as the quantum Duffing oscillator. The local maximum is in general stable.
However, in presence of resonant coherent or dissipative tunneling transitions
the system can escape from it and a stationary
state arises as a statistical mixture of quasienergy states being localized
in the two basins of attraction. This gives rise to a resonant or an
antiresonant nonlinear
response of the cavity at multiphoton transitions. The model
finds direct application in recent experiments with a driven superconducting
circuit QED setup.

\end{abstract}


\maketitle

\section{Introduction}
Damped nonlinear classical oscillators 
can display rather nontrivial features when they are modulated 
by an external time-dependent driving force 
\cite{Nayfeh}.  The simplest 
example is the classical 
Duffing oscillator, for which a quartic nonlinearity 
extends the harmonic potential such that the static potential still 
remains monostable. Adding a periodic modulation generates several 
dynamical stable states and the role of environmental fluctuating 
forces becomes particularly intriguing at the bifurcation points 
\cite{Dykman80}. Close to the fundamental resonance, the classical 
Duffing oscillator displays two stable states of large and small oscillation 
amplitudes. Environmental fluctuations can induce transitions between 
the two states and the scaling property of the probability of an activated 
escape from such a metastable state near the bifurcation point 
can be determined \cite{Dykman79,Dykman90}. 

For the corresponding quantum Duffing oscillator, two inherently quantum
mechanical effects 
can contribute to the escape from the metastable forced state over the 
effective quasienergy barrier. One mechanism 
is dynamical tunneling \cite{Dmitriev86} in the quasienergy surface, 
which leads at low temperatures to a sharp increase 
of the transition probabilities near the classical bifurcation point (determined
in the  quasiclassical approximation). In addition, quantum activation
\cite{Dykman88}
occurs which even at zero temperature leads to an effective diffusion over the 
quasienergy barrier induced by the environmental quantum fluctuations. 
At a first instance, this effect might appear somewhat counterintuitive,
but one has to bear in mind that energy absorption 
in the environment can increase the system's quasienergy, 
this quantity being defined in a rotating frame. 
It has been shown that the escape is always over the barrier (of activation
type), provided that 
the broadening of the quasienergy levels, induced by the environment, exceeds
the corresponding 
coherent tunneling rate\cite{Dykman88}. Other driven nonlinear quantum systems,
such as the parametrically 
driven Duffing oscillator \cite{Dykman06} or modulated large-spin
systems \cite{Dykman07} show quantum activated behavior as well. In this regime,
there is a separation of time scales. After the fast intrawell relaxation
processes have occured, the system occupies a quasiclassical 
state, which is metastable. From this, it can escape 
by rare interwell transitions that eventually would lead to the 
other metastable state. 

In the opposite limit, when tunneling transitions
noticeably contribute, the separation  of time scales is not well defined and a
mixture of all quasienergy states forms the stationary state. 
This is the case when a coherent resonant excitation  induces a population of a
multiphoton quasienergy state, as shown for the Duffing oscillator
\cite{Peano1,Peano2,Peano3}. Since the quasienergy states oscillate with 
different phases  with respect to the external 
modulation, a resonant or antiresonant response of the oscillator to the
modulation may occur at a multiphoton transition. Which type arises, 
is determined by the dominant stationary population of the involved quasienergy
state and thus 
depends also on the parameters of the environment. These lineshape properties 
around a multiphoton (anti-)resonance are connected 
with a resonantly enhanced escape in form of resonant dynamical tunneling 
\cite{Peano1,Peano2}, which  shows up as resonant tunneling peak in the
switching rate. 
Enhanced peaks are associated with resonant line shapes, while reduced peaks 
go with an antiresonance.

Recently, we have shown \cite{Peano10} that a dynamical bistability also occurs
in the 
setup of a driven linear resonator coupled to a quantum two-level system 
(driven circuit quantum electrodynamics (QED) set-up).  The system is 
conveniently modeled by the  Jaynes-Cummings (JC) model \cite{jaynescummings}
which is 
extended by a driving term. The static Jaynes-Cummings model 
 was originally studied to describe the interaction of a two-level atom and a
single 
quantized electromagnetic field mode and has an inherent 
nonlinearity since the splitting of the vacuum Rabi resonance depends 
on the number $N$ of photons in the resonator as $\sqrt{N+1}$. 
Already its undriven dynamics has many interesting facets, 
including Rabi oscillations, collapses and revivals of quantum states, quantum  
squeezing, quantum entanglement, Schr\"odinger cat and Fock states, and photon 
antibunching  \cite{Larson}. It also finds applications 
beyond quantum optical set-ups, namely in nanocircuit architectures, such as 
Cooper pair boxes \cite{Wallraff}, superconducting 
flux qubits \cite{Chiorescu}, Josephson junctions \cite{Hatakenaka}, 
and semiconductor quantum dots \cite{Winger}.  
In particular, the latter setups allow to explore the regime of 
strong qubit-resonator coupling as the resonator is typically formed 
by a nanoscale on-chip transmission line. 
In addition, strong driving allows to access the regime of nonlinear
response. 

These experiments based on quantum state engineering with superconducting 
circuits were accompanied by progress in theory, based on 
the adoption of the undriven JC model \cite{Blais04,Moon05,Wallquist06} 
to the particular experimental situations. In addition, the model has been 
extended by an additional time-dependent modulation term
\cite{Ficek02,Hauss08}, 
mimicking the effect of rf fields applied either to the qubit or to the
oscillator part. The more general case of $N$ two-level atoms strongly coupled to 
a driven optical cavity has been considered in Ref.\ \onlinecite{Xiao} 
for the case of weak driving fields and non-classical features in the photon 
correlation function have been identified.

The driven Jaynes-Cummings model has recently found an experimental realization 
in form of a superconducting transmon qubit device \cite{Bishop}. 
The transmitted heterodyne signal has been used as a measure 
of the amplitude of the stationary oscillations of the oscillator position. 
It has allowed to study for the first time the supersplitting of 
each vacuum Rabi peak which reflects transitions
between the groundstate and the first/the second excited state of the 
undriven Jaynes-Cummings spectrum.
Their energy difference is $2\hbar g$, where $g$ 
is the interaction strength of the qubit and the harmonic mode.  
In addition, for stronger driving, the excitation 
of discrete multiphoton transitions up to a photon number of $N=5$ has been 
observed. The measurements have been analyzed by accurate numerical 
simulations \cite{Bishop}. An  effective
 two-level approach which involves the vacuum and the one-photon dressed state
has 
been used to describe the vacuum Rabi splitting  \cite{Bishop,Carmichael}. 
 Moreover, the characteristic $\sqrt{N}$ spacing of the involved energy 
levels has been demonstrated.  At resonance, the system is coherently 
excited to a $N-$photon state. The relaxation occurs via subsequent 
dissipative transitions, generating eventually a steady state which 
involves a mixture of many quantum states and which renders a two-level 
description  inappropriate. 

In this paper, we complement our brief account of Ref.\ \onlinecite{Peano10} by
a 
comprehensive theoretical analysis of the driven Jaynes-Cummings model in the 
deep quantum regime of few photons in the resonator. Motivated by the failure of
perturbation theory for arbitrary small intensity 
of the driving, we carry on an alternative  approach.
The 
transformation to the frame rotating with the frequency of the external driving 
yields 
to a description in terms of quasienergy levels and states. Discrete
multiphoton 
transitions arise at avoided quasisenergy level crossings. Furthermore, we
obtain 
the Hamiltonian in the dressed state basis in form of two quasienergy surfaces,
one  of 
which shows bistability. This description allows to interpret the multiphoton
transitions as tunneling transitions
in the bistable quasienergy surface.  On the other hand, the region of the
spectrum for which perturbation theory fails can be studied
by means of a harmonic expansion of the quasienergy surface around its global
minimum.
 As already mentioned in Ref.\ \onlinecite{Peano10}, the lowest
quasienergy 
state is a quantum squeezed state with the squeezing parameter being a function
of the ratio 
of qubit-oscillator coupling and driving strength. In turn, this state of lowest
quasienergy 
has sub-Poissonian statistics. To complete the picture, we present the analysis
of the corresponding 
dissipative dynamics for the cases when both the qubit and the oscillator are
damped. In presence of resonant multiphoton transitions, the interplay of 
tunneling and dissipation results in a stationary state which 
consists of a statistical mixture of several quasienergy 
 states. Together with the levels involved in the
respective multiphoton/tunneling transition, 
also the levels around the global minimum may have a large 
weight in the mixture. We show that the stationary occupation probability of the
state with lowest quasienergy governs the lineshape of the 
nonlinear response yielding to resonant or antiresonant characteristics. 
As opposed to other anharmonic oscillators, the activation out of the
global minimum is not dominantly triggered by quantum activation
\cite{Dykman88,Dykman06}, but instead by dissipative
spin-flips (i.e., transitions between the two quasienergy surfaces). For
Purcell-limited devices and away from a 
multiphoton transition, this is the slowest dissipative process and the
corresponding rate can be computed by diagonalizing the Liouville operator 
numerically. Finally, we discuss the generic features of damped
driven nonlinear oscillators in the deep quantum regime.

\section{Coherent dynamics in the strong coupling regime}

We consider a state-of-the-art circuit \cite{Wallraff,Bishop,Schoelkopf} or optical 
\cite{Thomson,Haroche} 
cavity QED set up in the
strong coupling regime. The cavity is driven by an
external periodically time-dependent field. We note that this naturally implies 
the coupling to external modes which intrinsically renders the atom-cavity
system an open quantum system. However, the strong coupling regime is
characterized by an atom-cavity coupling which dominates over all dissipative
processes. In this limit, intrinsically coherent phenomena play a crucial role
and the dissipative processes can be considered as small perturbation. Hence, 
in order to lay the basis for the weakly damped, dissipative dynamics, we first 
 focus on the coherent dynamics of the atom-cavity system. This 
 paves the way to the full analysis of the open system which will be
carried out in Section \ref{sec.dissdyn}. Hence, at this stage it is not
necessary to restrict ourself to a specific configuration, e.g., a one- vs.\ 
a two-sided cavity or an circuit vs.\ an optical cavity. 

We model the cavity as an harmonic oscillator with frequency $\omega_{\rm r}$
which is characterized by the 
the ladder operators $a$ and $a^\dagger$ and which is coupled with strength $g$
to a qubit, modeling an (artificial) atom with two only relevant quantum states,
 with equal resonant frequency. The qubit is described in terms of 
the Pauli operators   $\sigma_{j=x,y,z}$ and the oscillator is modulated by 
a (classical) time-dependent field with frequency $\omega_{\rm ex}$ and field
strength $f$.  The total Hamiltonian thus reads 
($\hbar=1$)
\begin{eqnarray}\label{ham}
 H&=& \omega_{\rm r} \left( a^\dagger a+\frac{\sigma_z}{2}+\frac{1}{2}\right) 
+g \left(a^\dagger +a\right)\sigma_x \nonumber \\
& &  + f \left(a^\dagger+a\right) \cos \omega_{\rm ex} t \,. 
\end{eqnarray}
 In the frame rotating with $\omega_{\rm ex}$ and 
for the detuning $\delta \omega \equiv \omega_{\rm r}-\omega_{\rm ex}, g, f\ll
\omega_{\rm
r}$, we perform 
 a rotating-wave approximation (RWA) \cite{note} and obtain the Hamiltonian of
the driven JC model  as 
\begin{equation}\label{hamrwa}
 H=\delta \omega \left( a^\dagger a+\frac{1}{4}\sigma_+\sigma_-\right)
+\frac{g}{2} \left(a^\dagger \sigma_- +a\sigma_+\right)
  + \frac{f}{2} \left(a^\dagger+a\right) \, ,
\end{equation}
with $\sigma_\pm=\sigma_x\pm i\sigma_y$. 
Formally, the undriven JC model has the quasienergies ($n=1,2, \dots$)
\begin{equation}\varepsilon_{0}=0\,,\quad \varepsilon_{n,
\pm}=n\delta 
\omega \pm g\sqrt{n} \, , \label{dsquasienergies}
\end{equation}
which follow from diagonalization of the Hamiltonian in Eq.\ (\ref{hamrwa}) for
$f=0$. 
The quasienergy  states can be expressed in the product basis 
of oscillator eigenstates $\{|n\rangle\}$ and qubit 
eigenstates $\{|g\rangle,|e\rangle \}$ as 
\begin{eqnarray}|\phi_0\rangle&=&|0, g\rangle,\quad |\phi_{n\, -}\rangle
=\frac{1}{\sqrt{2}}(|n-1, e\rangle-|n, g\rangle)\nonumber\\
|\phi_{n\, +}\rangle&=&\frac{1}{\sqrt{2}}(|n-1, e\rangle+|n,
g\rangle).\label{dressedstates}
\end{eqnarray}
 We will refer to the two latter as dressed $n$-photon or Fock states with two
spin directions $\pm$. 
The $0-$photon quasienergy level crosses the $N$-photon level at 
\begin{equation}\label{resdet}
 \delta\omega=\pm g/\sqrt{N}
\end{equation}
(for spin $\mp$ respectively). 
For $f\ne 0$, the crossings turn into avoided crossings. At such an
anticrossing, 
 the zero-photon state $|\phi_0\rangle$ 
and the $N$-photon dressed state $|\phi_{N\pm}\rangle$ display Rabi oscillations
with 
the Rabi frequency $\Omega_N$ given by the minimal splitting of the two
quasienergy levels. 
These Rabi oscillations represent the $N$-photon transitions.  
 In order to have  well separated resonances, we consider here the strong
coupling regime 
$g\gg f$. There, one would naively expect that standard perturbation theory in
the driving $f$ 
 (and its generalization to quasidegenerate levels at resonance)
 would yield a correct description 
of all the quasienergy states which are relevant in the low energy dissipative
dynamics considered in Section \ref{sec.dissdyn}. 
However, in the remainder of this section we will show that for   $\delta\omega$
of the order
$\sqrt{fg}$ standard perturbation theory with respect to $f$ fails to describe
the lowest quasienergy part 
of the spectrum. In Section \ref{sec.dissdyn}, we will show
 that in fact {\em all\/} the quasienergy states play an important role in 
the dissipative dynamics close to a multiphoton resonance in the stationary
state. 

\begin{figure}[t]
\begin{center}
\includegraphics[width=60mm,keepaspectratio=true]{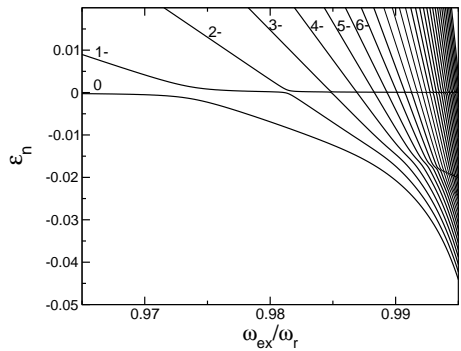}
\caption{\label{fig.1}Quasienergy spectrum (in unit of $\omega_{\rm r}$) of the
driven 
Jaynes-Cummings-Hamiltonian,  
Eq.\ \ref{hamrwa},  for the parameters $g=0.026 \omega_{\rm r},
f=0.004\omega_{\rm r}$.  
}
\end{center}
\end{figure}

 A first illustration of the breakdown of perturbation theory can be obtained by
numerically diagonalizing 
the  Hamiltonian 
in Eq.\ (\ref{hamrwa}). We use the Hilbert space spanned
 by the first $100$ dressed Fock states for each spin orientation.  The result
for the quasienergy spectrum 
for $g=0.026\omega_{\rm r}$ and $f=0.004 \omega_{\rm r}$ is shown in Fig.\
\ref{fig.1}. 
We note that we focus on positive detuning $\delta\omega>0$, 
since the opposite case trivially follows from 
$\delta\omega\to-\delta\omega$ and  
$|\phi_{n\pm} \rangle \to \exp[-i\pi a^\dagger a]|\phi_{n\pm} \rangle$ yielding 
$H\to -H$.
From Eq.\ (\ref{dsquasienergies}) it is clear that a signature of the validity
of 
standard perturbation theory in the driving amplitude $f$ is a constant slope of
the quasienergy
 levels $\varepsilon_{n, \pm}$ as a function of $\delta \omega$. 
When the quasienergy splitting between two states is small, i.e., of the order
of $f$, 
the two levels are to be considered quasidegenerate and
 an avoided crossing might occur. We clearly observe these patterns  almost
everywhere in the spectrum, 
even for levels corresponding to a large photon number
 that would appear in the region of large quasienergies not shown in Fig.\
\ref{fig.1}. The
notable exception is the region of small quasienergy and small detuning. There, 
 the slope of the levels with small quasienergies is not constant and no clear
avoided level crossing appears.
This clearly points to a
breakdown 
of perturbation theory in $f$ and it is easy to carry out a 
heuristic argument for this feature as follows.

Let us consider $\varepsilon_{n-}$  as a continuous function of $n$, which is
zero for $n=0$. It has a minimum 
$\varepsilon_{\rm min}=-g^2/(2\delta\omega)$ at $n_{\rm
min}=g^2/(4\delta\omega^2)$ and 
vanishes again for $n=g^2/\delta\omega^2$. Moreover, $g^2/\delta\omega^2$ can be
regarded 
as the number of states in the quasienergy interval $\varepsilon_{\rm
min}<\varepsilon<0$. 
The average quasienergy spacing $\bar{\varepsilon}$ in this interval  is thus
$\delta\omega/2$.
 Moreover it is clear that the levels tend to accumulate close to the minimum
and that $\bar{\varepsilon}$ overestimates the local level spacing there. The
matrix elements of the 
driving term in this region are of the order
 $\sqrt{n_{\rm min}}f$.
 We can conclude that the driving term is no longer a perturbation in the lowest
energy part of the spectrum if 
$\sqrt{n_{\rm min}}f\sim\bar{\varepsilon}$ or  $\delta\omega\sim\sqrt{fg}$. 
This rough estimate is confirmed by the numerical data. In fact, for $f=0.004
\omega_{\rm r}$ and $g=0.026 \omega_{\rm r}$
as in Fig.\ \ref{fig.1}, we find $\sqrt{fg}\approx0.01$, which correspond
approximately to the region where the slope of the levels in not constant.

\section{Quasienergy landscape and dynamical bistability}

Motivated by the failure of ordinary perturbation theory, we next 
formulate a different perturbative approach. 
For this, it is convenient to switch to the picture of dressed qubit states,
formally
achieved by the unitary transformation \cite{persico}
\begin{equation}
 R=\exp\left[{\frac{-3\pi}{8\sqrt{a^\dagger a+\sigma_+\sigma_-/4}}(a^\dagger
\sigma_--a\sigma_+)}\right] \, .
\end{equation}
It maps the JC eigenstates Eq.\ (\ref{dressedstates}) into product states
according to 
$$|0, g\rangle\to|0, g\rangle,\quad |\phi_{n-}\rangle
\to|n, g \rangle,\quad |\phi_{n+}\rangle\to-|n-1, e\rangle.$$
 The purpose of this transformation is that 
the undriven JC Hamiltonian becomes diagonal in the qubit Hibert space and
assumes the form 
\begin{eqnarray} \label{htilde}
\tilde{H}=|\delta \omega | \left( a^\dagger a+\frac{1}{4}\sigma_+\sigma_-\right)
 + g \sigma_z \sqrt{a^\dagger
a+\frac{1}{4}\sigma_+\sigma_-} \, ,
\end{eqnarray}
while the ladder operators remain unaffected to lowest order in the photon
number, i.e., 
\begin{equation}
\tilde{a}=R^\dagger a R = a+{\cal O}(n^{-1/2})\,.\label{arotated}
\end{equation}
 The expression for $\tilde{a}$ follows by expanding the matrix elements
\begin{eqnarray}
\label{fplusminus}
 \langle ng|\tilde{a}|mg\rangle &=& 
\langle ng|R^\dagger a R |mg\rangle  = 
\langle \phi_{n-}|a|\phi_{m-}\rangle \nonumber \\ 
& = & 
f_+(n)\delta_{n+1 m} \, ,\nonumber\\
\langle ng|\tilde{a}|me\rangle &=& 
\langle ng|R^\dagger a R |me\rangle  = 
\langle \phi_{n-}|a|\phi_{m+}\rangle \nonumber \\ 
& = & 
f_-(n)\delta_{n+1 m}\,,\label{dsamat}
\end{eqnarray}
with $f_\pm(n)=(\sqrt{n}\pm\sqrt{n+1})/2$. 
Note that $f_-(n)$ is typically small, e.g., for $n=1$ (it is the largest case),
$f_-(1)\approx -0.2$ and it approaches zero as $-1/(4\sqrt{n})$. 
On the other hand, $f_+(n)\sim \sqrt{n}+{\cal O}(1/\sqrt{n})$. 
This also illustrates that only higher order terms depend on spin flipping
operators. 
Hence, it follows that both the driving and the coupling to the bath induce 
spin flips, but only as higher order processes. 

 Next, we introduce the rescaled rotating quadrature 
\begin{equation}\label{resquad}
 {\cal X}=\sqrt{\frac{\lambda}{2}}(a^\dagger+a)\,,\quad\quad{\cal
P}=i\sqrt{\frac{\lambda}{2}}(a^\dagger-a),
\end{equation}
 with
$\lambda=|\delta \omega|^2 / g^2$.   
Note that the terms neglected in Eq.\ (\ref{arotated}) are of
higher order in $\lambda$. 
By plugging the rotating quadrature into the transformed Hamiltonian
Eq.\ (\ref{htilde})  and neglecting all higher order terms, 
we obtain  $\tilde{H}\simeq \delta\omega\lambda^{-1} Q ({\cal X}, {\cal P})$
with 
\begin{equation} \label{hampotfull}
Q ({\cal X}, {\cal P})  =   \frac{{\cal X}^2}{2}+\frac{{\cal P}^2}{2} + \sigma_z
 \sqrt{\frac{{\cal X}^2}{2}+\frac{{\cal P}^2}{2}} 
+ \frac{f}{\sqrt{2} g} {\cal X}   \, .
\end{equation} 
Eventually, rescaling the time by $\tau=1/\delta\omega$, yields the effective 
Schr\"{o}dinger equation 
 \begin{equation}
i \lambda\frac{\partial \psi}{\partial t}\simeq  Q ({\cal X}, {\cal
P})\psi\,.
 \end{equation}
with the commutator $[{\cal X}, {\cal P}]=i\lambda$ (which follows from the
definition in Eq.\ (\ref{resquad})).
\begin{figure}
\begin{center}
\includegraphics[width=85mm,keepaspectratio=true]{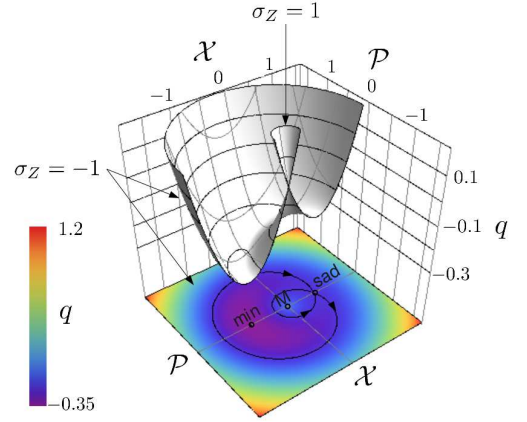}
\caption{\label{fig.2} 3D  plot of the quasienergy surfaces $Q({\cal X},
{\cal P})$  for ${\cal P}>0$ and   
$f/g=0.154$ (Note that this yields an effective temperature of  $T_{\rm
eff}=0.65$, see text). The quasienergy surface for the dressed spin orientation
$e$ ($\sigma_z=+1$) is a monostable  reversed cone. 
The one for the opposite dressed spin $g$ ($\sigma_z=-1$) is bistable and is
shown as a density plot as well.
In the density plot, the oriented black solid line indicates
the separatrix between the two domains of attraction.}
\end{center}
\end{figure}
 Thus,  we can interpret $\lambda$ as a rescaled effective Planck constant,
${\cal X}$
and ${\cal P}$ as canonically conjugated operators and $Q ({\cal X}, {\cal P})$ 
as two quasienergy surfaces in phase space for the two opposite dressed spin
orientations. 
They are  visualized in Fig.~\ref{fig.2}. 
The quasienergy surface for the dressed spin $g$ ($\sigma_z=-1$)  has the
overall shape of a Mexican
hat. The drive induces a finite tilt of the surface, generating a 
saddle point at ${\cal P}_{\rm sad}=0$ and 
${\cal X}_{\rm sad}=(1-f/g)/\sqrt{2}$ and
a separatrix, shown in the density plot in Fig.~\ref{fig.2}  as the oriented
black solid line. It
divides the surface into three domains: i) a potential well around the 
quasienergy minimum at ${\cal P}_{\rm min}=0$ and ${\cal X}_{\rm
min}=-(1+f/g)/\sqrt{2}$, ii) an internal dome around 
the inner maximum at  ${\cal X} = 0,{\cal P}=0$ and, iii) an external surface.
For quasienergies lying above the saddle point and below the maximum, there are 
orbits coexisting on the domains ii) and iii) and  
thereby define a dynamical bistability. The surface for 
the opposite dressed spin orientation $\sigma_z=+1$  is a less
 interesting monotonous function and is shown in the 3D plot in
Fig.~\ref{fig.2}. 

\begin{figure}[t]
\begin{center}
\includegraphics[width=72mm,keepaspectratio=true]{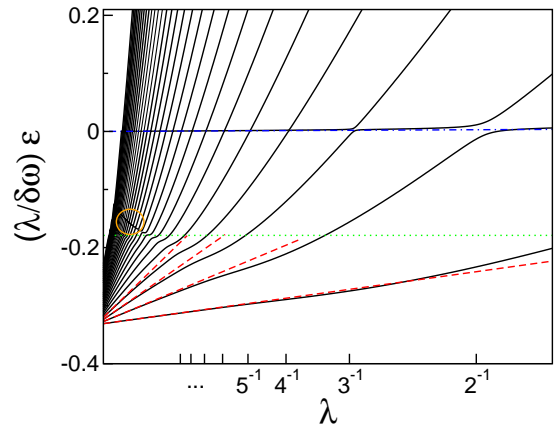}
\caption{\label{fig.3} Rescaled quasienergy spectrum of the driven
Jaynes-Cummings Hamiltonian in 
Eq.\ (\ref{hamrwa})  for the parameters $f/g=0.154$ (solid black lines). The
green dotted line indicates the saddle point quasienergy. 
 The red dashed lines correspond to the approximate $(\lambda/\delta\omega)
E_n$, see text. The blue dotted-dashed  line marks  the rescaled quasienergy
$\varepsilon_0$ of $|0(f)\rangle$. The orange circle highlights  avoided
quasienergy level crossings (multiphoton transitions) for large photon numbers
$N$. }
\end{center}
\end{figure}

When the motion is quantized, the (quasi-)energy levels become discrete. For
small $\lambda$, we expect the dynamical behaviour to be semiclassical
and we can associate each quasienergy level to an allowed orbit. In principle, a
full
WKB treatment is possible but it goes beyond the scope of this paper.
In order to illustrate the semiclassical features of the driven Jaynes-Cummings
model, we show the
quasienergy levels (rescaled by $\lambda/\delta\omega$) as a function of
$\lambda$ in Fig. \ref{fig.3}. They are obtained by numerically diagonalizing
the complete Jaynes-Cummings Hamiltonian. 

We can associate the levels below the quasienergy saddle point $Q({\cal X}_{\rm
sad},0)=-(1-f/g)^2/4$ (indicated in Fig. \ref{fig.3} as a green dotted line)
to orbits localized in the quasipotential well. Notice that close to the minimum
$Q({\cal X}_{\rm min},0)=-(1+f/g)^2/4$, the orbits are approximately
equally spaced and they become denser while approaching the saddle point. In
fact, semiclassically the level spacing is given by $\lambda \omega(q)$, where 
$\omega(q)$ is the 
classical orbit frequency for the (rescaled) quasienergy $q$. Close to the
quasienergy minimum the potential is harmonic and  $\omega(q)$ varies slowly
while close to the saddle point, it tends quickly to zero. 

For energies above the saddle point and below the maximum, we can associate the
quasienergy levels with orbits in the internal dome
and in the external surface. The orbit with vanishing quasienergy corresponds to
an orbit close to the quasienergy maximum.  For certain discrete values of
$\lambda$,  an orbit on the external surface might have
vanishing quasienergy as well. Hence, we can interpret the multiphoton
transitions, which occur at the resulting avoided
level crossings, as tunneling transitions
between two states in the internal dome and the external well, respectively. In
the limit of $g\gg f$, we can read off the resonant tunneling condition 
from the perturbative result in  Eq.\ (\ref{resdet}), yielding the condition 
for $N$-photon transitions to be 
$\lambda=1/N$. More tunneling transitions for negative quasienergy are
indicated by the orange circle in Fig. \ref{fig.3}.

Above the minimum no avoided level crossing is present, several exact crossings
(not shown) appear instead. The corresponding orbits have opposite dressed spin
and do not show level repulsion because the two quasienergy surfaces are
effectively decoupled.

\subsection{Quasienergy states at the well bottom}

Next, we study the spectrum close to the bottom of the quasienergy well
quantitatively. 
In fact, as it turns out, the states localized in this region 
play an important role in the stationary dissipative dynamics 
close to a multiphoton transition.

The most simple approach  consists
in  expanding the quasienergy in Eq.\ (\ref{hampotfull}) close to its
minimum. 
In this region,  the quasienergy surface is to lowest order 
harmonic and follows
\begin{eqnarray}
 Q({\cal X},{\cal P})&=&Q({\cal X}_{\rm min},{\cal P}_{\rm min}) 
+ \frac{1}{2m_{\rm eff}} ({\cal P}-{\cal P}_{\rm min})^2 \nonumber \\
 & &  + \frac{1}{2} m_{\rm eff}{\omega^*}^2 ({\cal X}-{\cal X}_{\rm min})^2 \, ,
\end{eqnarray}
with effective mass $m_{\rm eff}=(f+g)/f$ and 
with frequency
$\omega^* = \sqrt{f/(f+g)}$. When the zero point energy $\lambda\omega^*/2$ is
much smaller than the quasienergy well depth 
$\Delta Q\equiv Q({\cal X}_{sad},{\cal P}_{sad})-Q({\cal X}_{min},{\cal
P}_{min})=f/g$,  only a   
few quasienergy states are localized in the well.  
The harmonic expansion  
yields the quasienergies   
\begin{equation}
 E_n = \delta\omega\lambda^{-1} \left[Q({\cal X}_{min},0) +\lambda
\omega^*\left(n+\frac{1}{2}\right)\right]
\label{harmen}\,,
\end{equation}
which are determined up to
${\cal O}(\lambda^2)$ \cite{note2}. They are shown in Fig.\ \ref{fig.3}  as red
dashed lines and they
almost coincide with the exact results for small $\lambda$.

Up to leading order in $\lambda$, the corresponding wave functions are given by
\begin{equation}|\psi^*_n\rangle=R^{-1}\frac{b^{\dagger n}}{\sqrt{n!}} D({\cal
X}_{min})S(r^*)|0\, 
g\rangle\label{lowestenergystate2} \, .
\end{equation}
They are obtained by applying to the vacuum:
i) the squeezing operator $S(r^*)=\exp \left[r^*(a^{ 2}-a^{ \dagger2})/2\right]$
with squeeze factor  
\begin{equation}\label{squeezingfactor}
r^*=\frac{\ln [m_{\rm eff}\omega^*]}{2}=\frac{\ln [1+g/f]}{4}\,,
\end{equation}
ii) the translation $D({\cal X}_{\rm min})=\exp \left[i {\cal P} {\cal X}_{\rm
min}/\lambda\right]$   to the  minimum, 
iii) the creation operator $b^\dagger= a^\dagger\cosh{r}+ a\sinh{r}-e^{r}{\cal
X}_{\rm min}/\sqrt{2\lambda}$, 
 and iv) $R^{-1}$ to return to the bare qubit picture. With this, one can
compute
 the Fock-state representation \cite{yuen} of the quasienergy states and all
expectation values at leading order.  

\subsection{Quantum squeezed state and sub-Poissonian statistics}

Of particular interest is the state $|\psi^*_0\rangle$ at the bottom of the
quasienergy well.
It is defined, when  the zero point energy $\lambda\omega^*/2$ is smaller than
the quasienergy depth 
$\Delta Q=f/g$,
 corresponding to the region $\delta\omega<g\sqrt[4]{f/g}$ ($=0.016$ for the
parameters used in 
Fig.\ \ref{fig.1}). 
Note that the analytical result for $E_0$ in Eq.\ (\ref{harmen}) almost
coincides
 with the exact one in this region. Most importantly, $|\psi^*_0\rangle$ 
exists for any finite driving (providing that the detuning is small enough), 
and is clearly absent in the undriven case. In fact, for $f=0$, the tilt of 
the quasienergy surface vanishes and
 no quasienergy well is developed. This illustrates further that simple 
perturbation theory  fails for arbitrary small driving close to zero detuning
and 
improves the estimate of the critical detuning below which it breaks down.

For weak but finite driving, the well is still very shallow in the momentum
direction, 
allowing for large momentum fluctuations of a state confined in it. For
increasing driving, the well becomes deeper 
and more symmetric.  Hence,$|\psi^*_0\rangle$ is amplitude squeezed and the
squeezing decreases for increasing 
driving, as it can also be read off from Eqs.\ (\ref{lowestenergystate2}) and
(\ref{squeezingfactor}).

As a consequence,  $|\psi^*_0\rangle$  has also sub-Poissonian statistics and
shows photon
antibunching. This can be seen by computing
 its average photon number 
$\bar{n}^*\equiv\langle\psi^*_0|n|\psi^*_0\rangle$ and its variance $(\Delta
n^*)^2\equiv\langle\psi^*_0|(n-\bar{n}^*)^2|\psi^*_0\rangle$ 
by means of Eq.\ (\ref{lowestenergystate2}). We find 
\begin{equation}
 \bar{n}^*=\frac{{\cal X}_{\rm min}^2}{2\lambda}=\frac{(f +
g)^2}{4\delta\omega^2}\, ,\label{nmed}
\end{equation}
 and 
\begin{equation}
(\Delta n^*)^2= e^{-2r^*}\bar{n}^*=e^{-2r^*}\frac{(f +
g)^2}{4\delta\omega^2}   \, .
\end{equation}
Thus, in the semiclassical limit $\lambda \to 0$, the $2$-photon correlation
function becomes
\begin{equation}
 g^{(2)}(0)\equiv1+\frac{(\Delta n^*)^2-\bar{n}^*}{\bar{n}^{*2}}<1\,,
\end{equation}
Note that the above formulas can be regarded only as leading-order estimates,
because
they are 
of order $1/\lambda$ and other finite contributions are not taken into
account. 

Here, we did not include the Fock representation of the semiclassical states
because it is rather cumbersome
and scarcely illuminating for the purpose of this work. 
However, it is important to keep in mind that the semiclassical states are a
superposition of dressed 
Fock states with a given dressed spin orientation and photon number $n$ of the
order of 
$\bar{n}$, given in Eq.\ (\ref{nmed}). For the discussion of the dissipative
dynamics below, we remark that close to a $N$-photon transition, all the dressed
Fock  states that contribute significantly to the superposition correspond a
photon number $n<N$ with negative 
 quasienergies, $\varepsilon_{n<N-}<0$.

Finally, we note that the Golden Rule estimate for the rate for coherent spin
flip transitions yields zero, because no  
overlap between the quasienergy states localized in the well (which has negative
quasienergy $E_n<0$) and 
that with opposite spin (which has positive quasienergy $\varepsilon_{n+}>0$)
exists. 

In this work we consider identical resonant frequencies for the qubit and the
cavity. However close to the quasienergy minimum the effect of a finite detuning
$\delta=\omega_{\rm q}-\omega_{\rm c}$ of the resonant frequencies $\omega_{\rm
q}$ and  
$\omega_{\rm c}$ for the qubit and the cavity might be  negligible even if the
circuit cavity QED is in the dispersive regime when few photons are present.
The detuning $\delta$ induce an additional term $\delta^2\lambda/(4 g^2)$ inside
the square root in Eq.\ (\ref{hampotfull}).
This further contibution is negligible close to the well bottom, if
$\delta^2\lambda/ g^2\ll 2 {\cal X}_{\rm min}^2=(1+f/g)^2$. Note that this
condition can be fullfilled even if $\delta\gg f,g$ as in Refs.\
\cite{Scholkopf,Girvin}.  
In this references, the state $|\psi^*_0\rangle$ has been denoted as bright
state.

\subsection{Dark state and multiphoton transitions}

Another state playing a key role in the dissipative dynamics is the state
obtained by starting in the JC groundstate $|0,g\rangle$ and by adiabatically 
switching on 
the driving. This is the state with zero quasienergy and is shown as blue
dotted-dashed  line in Fig.\ \ref{fig.3}. We denote it as $|0(f)\rangle$. It is
naturally favored by photon leaking,
the most important dissipation channel in many set-ups. For weak
driving, it has vanishing average photon number and therefore it is not
accurately described by our semiclassical approach.

This state has been investigated in Ref.~\onlinecite{Milburn}  for
$\delta\omega=0$. It is a squeezed state which tends to follow the
driving by rotating its spin by the angle $\theta\equiv\arcsin f/g$
around the  the $y-$axis and squeezing its amplitude fluctuations with squeezing
factor $r_0=-[\ln (1-f^2/g^2)]/4$ \cite{Milburn}.
 Note that, as opposed to $r^*$ (the squeezing factor of $|\psi^*\rangle$), 
$r_0$
{\em increases\/} for increasing driving. In addition, it has super-Poissonian
statistics and shows photon bunching. 
In fact, 
\begin{eqnarray}
 \bar{n}_0&\equiv&\langle 0 (f)|n|0 (f)\rangle_f=\sinh^2 r_0\, ,\label{nmed0}\\
(\Delta n_0)^2&\equiv&\langle 0(f)|(n-\bar{n}_0)^2|0(f)\rangle=2 \cosh^2 r_0
\sinh^2 r_0  \, .
\end{eqnarray}
and the $2$-photon correlation
function is
\begin{equation}
 g^{(2)}(0)\equiv1+\frac{(\Delta n)^2-\bar{n}}{\bar{n}^2}>1\,.
\end{equation}
Since it is not possible to diagonalize analytically the driven JC 
Hamiltonian for $\delta\omega=0$, a systematic perturbation theory in
$\delta\omega$ is not possible. Fortunately, for weak driving, we can rely on 
perturbation theory in the driving $f$ to study this state. Away
from a multiphoton transition, we can compute the transmission  $\langle
0(f)|a|0(f)\rangle$ by ordinary perturbation theory in $f$, yielding
\begin{equation}\label{a00}
 \langle 0(f)|a|0(f)\rangle\simeq -\frac{f}{4}\left(\frac{1}{\delta \omega+g}+
 \frac{1}{\delta \omega-g}\right) \, . 
\end{equation}
Hence, the oscillation of the transmission is in phase with the driving for
$\delta\omega<g$. 

At a multiphoton transition, i.e, for $\lambda\approx 1/N$, the corresponding
multiphoton state is a superposition of $|0(f)\rangle$ and a 
semiclassical state $|N(f)\rangle$, which exists on the external part of the
quasienergy surface. 
This state is obtained by starting from a dressed state $|\phi_{N+}\rangle$
($|\phi_{N-}\rangle$ for $\delta\omega>0$) and switching on adiabatically the
driving.  The states
 $|0(f)\rangle$ and  $|N(f)\rangle$ display Rabi oscillations.
The corresponding Rabi frequencies can be computed by means of Van Vleck
perturbation theory as
\begin{equation}\label{Rabifreq}
 \Omega_1=\frac{f}{\sqrt{2}},\,\,\Omega_2=\sqrt{2}\frac{f^2}{g},\,\,
\Omega_3=\frac{3^{7/2}}{2^4}\frac{f^3}{g^2}\,\cdots
\end{equation}
We do not give the general formula because it is rather cumbersome and scarcely
illuminating. In fact, it is derived assuming that perturbation theory  is valid
everywhere in the spectrum, 
which is generally not true, not even for $f\ll g$. Our  approach does not allow
a fully-fledged semiclassical calculation of the Rabi frequency. Nevertheless,
one main advantage is that it describes very elegantly how the physical
quantities are rescaled while the effective Planck constant changes. With this,
we postulate that the Rabi frequency follows as 
\begin{equation}
 \Omega_N\sim \exp{\left[-S(f/g)/\lambda\right]}=\exp{\left[-N S(f/g)\right]}\,.
\end{equation}
with $S(f/g)$ being an unknown function of $f/g$ and $\sim$ indicating
logarithmic precision.
  
\section{Dissipative dynamics}
\label{sec.dissdyn}
  State-of-the-art nanocircuit QED set-ups  \cite{Wallraff,
Chiorescu,Winger,Bishop} 
are characterized by weak damping, implying large quality
factors of the order of $Q\sim 10^{4} - 10^{5}$ 
 and qubit dephasing and relaxation times ($T^*_1$ and
$T_2$) being large
compared to the timescale $T_r=2\pi/\omega_r$,  
governing the system dynamics. For instance, for the transmon 
architecture reported in Ref.\ \onlinecite{Bishop}, $T_{1,2} \sim 1 \mu$s and
 $T_r \sim 1$ ns. For optical cavities, quality factors of 
$Q\sim 10^{10}$ are possible \cite{Haroche}. 
When decoherence and dissipation are induced by electromagnetic environmental
fluctuations 
 with a smooth spectral density  and when all the time scales governing the
different 
dissipative processes exceed typical bath-intrinsic correlation times, a
Markovian dynamics is expected. 
The simplest Markovian master equation (MME) for the reduced density operator
$\rho(t)$ of the 
qubit-plus-oscillator system is of Lindblad form and incorporates oscillator
relaxation, 
qubit dephasing and qubit relaxation. Its standard form is given by
$\dot{\rho}={\cal L}[\rho]$, with the Liouvillian
\begin{eqnarray}
{\cal L}[\rho]&\equiv&-i[H,\rho]+\kappa{\cal D}[a]+\gamma_1{\cal
D}\left[\sigma_-/2\right]
+\frac{\gamma_\varphi}{2}{\cal D}\left[\sigma_z\right],\quad
\label{fullME}
\end{eqnarray}\\
where  ${\cal D}[O]$ is the Lindblad damping superoperator ${\cal
D}[O]\equiv([O\rho,O^\dagger]+[O,\rho O^\dagger])/2$. 
The three damping terms describe: 
i) the photon leaking out of the oscillator 
at rate $\kappa$, ii) intrinsic qubit 
relaxation at rate $\gamma_1$, and iii)  pure qubit dephasing at rate
$\gamma_\varphi$.  
This phenomenological master equation follows by modelling the environment as 
three independent harmonic baths, each 
  held at thermal equilibrium at the same 
temperature $T$,  provided that
$\hbar\kappa,\hbar\gamma_1,\hbar\gamma_\varphi\ll k_{\rm B} T, \hbar \omega_r$. 
In Eq.\ (\ref{fullME}), we have implicitly assumed a one-sided cavity
 with a single input and a single output port, yielding
 a single dissipative channel for the cavity. 
For a detailed description of a nanocircuit and an optical
set-up implementing this model, we refer the reader to 
 Ref.\ \onlinecite{Bishop} and Ref.\ \onlinecite{Thomson}, respectively.
 
Moreover, one has to assume $k_B T\ll\hbar\omega_r$. This condition together
with the ordinary assumptions for the RWA 
($\delta\omega, g, f\ll\omega_r$) implies that the environment acts as a perfect
energy sink. However,  quasienergy
 and not energy is the
good quantum number for  driven systems. Since the quasienergy is defined in a
rotating frame,  
energy emission in a static frame (energy leaking into the environment)
 can appear as quasienergy absorption in the rotating frame. 
This leads to counterintuitive effects, such as the Unruh effect for a
constantly accelerated relativistic system  \cite{Unruh} 
and  quantum activation for the quantum Duffing oscillator \cite{Dykman88}.
In the following, we will illustrate how our approach in terms of the
quasienergy surface provides an intuitive physical insight into the dissipative
dynamics even when many different quantum states are involved.

\subsection{Potential well escape: quantum activation vs. spin flips}
\label{secnonres}

Away from any multiphoton transition and for $g\gg f$, $|0(f)\rangle$ is
characterized by a vanishing average photon number. 
If the system acts as an energy sink, this state 
has an infinitely large lifetime because a photon cannot be emitted neither from
the system nor from the environment.
Hence,  $|0(f)\rangle$ will be dominantly populated in the stationary state. 
Conversely, at a multiphoton resonance, the system can escape from the
$0-$photon state 
$|0(f)\rangle$ via resonant tunneling to the dressed $N-$photon state
$|N(f)\rangle$. 
Subsequent relaxation causes the energy to leak out from the system via 
transitions to states with lower quasienergies. 
This leads to the occupation of 
all the quasienergy states with negative quasienergies. In fact, these states 
are a superposition of dressed Fock states with photon number $n<N$, as 
detailed 
in the previous section. Hence, when $\delta \omega<g\sqrt[4]{f/g}$, energy
leaking leads 
to the occupation of those quasienergy states which are 
localized {\em in\/} the quasienergy well. 

From the basin of attraction of the quasienergy minimum, the system can escape 
either i) by  climbing up the quasienergy well by quantum activation, or ii) by
means of a spin flip 
transition. In fact, only one 
of the two dressed spin states is confined. Hence, after a spin flip 
transition the system can quickly decay to the dark state. The rate
$\Gamma_a$ for quantum
activation  is suppressed exponentially with the number $N_{\rm w}$ of states in
the quasienergy well, following
$\Gamma_a\propto \kappa e^{-c N_{\rm w}}$. Note that the constant $c$
does depend on the precise shape
of the quasienergy surface, but not on the rescaled Planck constant $\lambda$,
whereas $N_{\rm w}$ obviously depends on $\lambda$ (which governs the
 level spacing) according to $N_{\rm w}\propto 1/\lambda$. Hence, spin flip
transitions are the dominant escape channel when $\lambda$ is small.

The stationary state is a statistical mixture of all those quasienergy 
states with negative quasienergy, if the overall rate of escape from the
quasipotential well is of the same order as the rate of escape from the
state $|0(f)\rangle$. The latter is discussed in the next subsection.  

\subsection{Coherent vs. incoherent resonant dynamical tunneling}
\label{sectunneling}
Driving may also induce resonant transitions between the qubit dressed states. 
 To be more precise, we have to distinguish between the regimes of
(A) coherent and (B) incoherent resonant dynamical tunneling. The borderline
between these 
two regimes is determined   
by the ratio $\Omega_N/\Gamma_N$, where $\Omega_N$ is the Rabi frequency of the 
$N-$photon transition and $\Gamma_N$ is the inverse lifetime of the dressed
$N-$photon state. The latter is of the order of the maximum of 
$N\kappa, \gamma_1, \gamma_\phi$. 

{\em (A) Coherent resonant dynamical tunneling\/} is characterized by the decay
rate of the dressed $N-$photon state being smaller than the Rabi
frequency, $\Gamma_N<\Omega_N$. Then, the system coherently tunnels back
and forth with the Rabi frequency $\Omega_N$ several times before it
significantly relaxes with decay rate $\Gamma_N/2$. 
All the subsequent dissipative transitions occur on a time scale
similar to $\Gamma_N$. Therefore, the stationary solution will be a statistical
mixture of all the states with negative quasienergies. This is qualitatively
different from the situation away from resonance where only the $0$-photon state
is significantly populated. 

{\em (B) Incoherent resonant dynamical tunneling\/} occurs when the decay rate
of the $N-$photon dressed state exceeds the Rabi frequency, $\Gamma_N>\Omega_N$.
Then, the quasienergy 
level broadening is larger than the coherent splitting of the two quasienergy
levels. 
The dressed $N-$photon state strongly fluctuates on a quasienergy range
$\Gamma_N$ and with it the quasienergy difference to the $0-$photon state. 
Tunneling occurs only for those splittings which do not exceed the Rabi
frequency $\Omega_N$. This occurs with probability  
$\Omega_N/\Gamma_N$. When the two levels are quasi-degenerate, a tunneling
event then happens with probability $\Omega_N$. 
For incoherent resonant dynamical tunneling, the total tunneling 
probabilty is just the product of the two. 
Hence, the system can escape from the $0-$photon state $|0\,g\rangle$ with a
small total rate $\Omega_N^2/\Gamma_N$.
In the incoherent regime, again two situations can arise: 

{\em (B1)} When no state is
localized in the quasienergy well (which roughly occurs 
for $\delta \omega>g\sqrt[4]{f/g}$, see above), i.e., when perturbation theory
in the driving $f$ is valid, the lifetimes of all the 
states visited during the relaxation transition are of the same order, namely
of the order of $\Gamma_N^{-1}$. 
This is given by the maximum of $\kappa^{-1},
\gamma_1^{-1}, \gamma_\phi^{-1}$. This,
however, is in any case much smaller than the lifetime of the $0$-photon state,
given by  $\Gamma_N/\Omega_N^2$ (remember that $\Gamma_N$ is suppressed
exponentially with $N$). Thus, the $0$-photon state is maximally populated. 

{\em (B2)} In the second case, when several states are trapped in the
quasienergy well
(which  occurs for $\delta \omega\ll g\sqrt[4]{f/g}$), i.e., 
when the semiclassical description is necessary, the rate for quantum activation
 $\Gamma_a$ is exponentially small and can be of the order of
$\Omega_N^2/\Gamma_N$. 
If also the spin flip rate is small, the state at the well bottom can become
metastable. In this case, it is therefore possible that an 
overall small rate of incoherent resonant dynamical tunneling leads to a 
dramatic change in the stationary distribution as compared to the situation 
away from resonance, resembling the situation of
coherent resonant dynamical tunneling. Note that, this will be the case only for
the first few resonances. In fact the resonant escape from the  dark state 
is suppressed exponentially with $N$, which is not the case for the spin flip
escape from the bright state. 

\section{Nonlinear response in the strong coupling regime} 
\label{sec.purcell}

Having discussed all required ingredients, we can next turn to 
 the nonlinear response of the driven JC model, which can be 
 characterized by the steady-state expectation value of the 
intracavity field operator related to 
\begin{equation} \langle a  \rangle = {\rm tr}(\varrho  a) = A e^{i\varphi} 
 = \sum_{\alpha \beta}\varrho_{\alpha\beta} a_{\beta\alpha}\, . 
\label{overalloscillation}
\end{equation}
The indices $\alpha,\beta$ refer to the basis of eigenstates of 
the driven JC Hamiltonian in Eq.\ (\ref{hamrwa}). 
In a one-sided cavity, its modulus $A$
is related to the transmitted amplitude 
$A_{\rm tr}\propto A$ and intensity $ I_{\rm tr}\propto A^2$ of the input signal 
in a heterodyne measurement set-up \cite{Bishop,Thomson}.

These quantities are directly accessible in the experiment. 
In the rotating frame, $\langle a\rangle< 0$ 
corresponds to an oscillation out of phase ($\varphi=\pi$) with respect to the
drive, while 
the in-phase oscillation ($\varphi=0$) is associated to $\langle a\rangle> 0$. 
Note that  $\delta\omega\to-\delta\omega$ and
$\varrho_{\alpha\beta}\to\varrho^*_{\alpha\beta}$, since 
the unitary transformation $\exp{[-i\pi a^\dagger a]}$ yields ${\cal
L}\to{\cal L}^*$ . Moreover, 
$a_{\beta\alpha}\to-a_{\beta\alpha}$ and all matrix elements are real. We can
conclude that the nonlinear response for 
$\delta\omega<0$ follows trivially from the one for  $\delta\omega>0$ according
to
$\langle a  \rangle\to-\langle a  \rangle^*$, i.e. $A\to A$ and $\varphi\to
\pi-\varphi$. In addition, due to the RWA, the master equation can be written in
terms of the ratios 
$f/g$, $\kappa/g$, $\gamma_1/g$ , $\gamma_\phi/g$ and $\lambda$ only. Thus, 
also $\langle a \rangle$ depends only on these quantities.
\begin{figure}[t]
\begin{center}
\includegraphics[width=60mm,keepaspectratio=true]{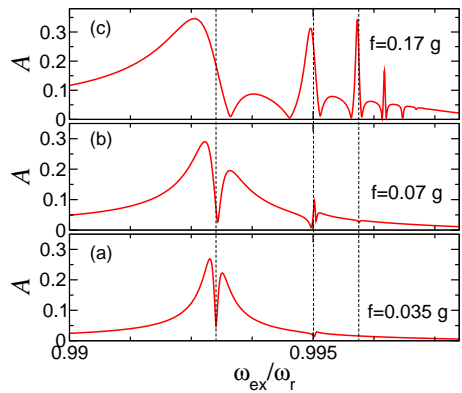}
\caption{\label{fig.4}Nonlinear response of the driven JC model: amplitude $A$
as a function of the driving frequency
$\omega_{ex}$ for $f=0.035 g $ (a), $f=0.071 g$ (b)
and $f=0.17 g$ (c).  Moreover, 
$\kappa=6.1 \times 10^{-3} g$, $\gamma_1=1.4 \times 10^{-3} g$,
$\gamma_\phi=1.4\times 10^{-4}g$, $g=0.007 \omega_{\rm r}$. 
A realistic value for $\omega_{\rm r}/(2\pi)$ is $7$ GHz \cite{Bishop,Houck}.  
}
\end{center}
\end{figure}

A straightforward numerical solution of the stationary limit $\dot{\rho}(t)=0$
allows to 
numerically calculate the modulus $A$ as a function of the external modulation 
frequency $\omega_{\rm ex}$. The result for experimentally realistic parameters 
\cite{Bishop} 
 is shown in Fig.\ \ref{fig.4}. For weak modulation,  two
large  
fundamental (anti)resonances appear symmetrically with respect to $\omega_{\rm
ex}=\omega_{\rm r}$, which mark the supersplitting of the 
vacuum Rabi resonance (in Fig.\ \ref{fig.4}a, we show only the regime of 
$\omega_{\rm ex}<\omega_{\rm r}$). 
These largest resonances (which are in fact
antiresonances) are associated to the $1$-photon transitions. 
They can be described \cite{Bishop,Peano3} by an effective two-level model 
involving the $0$- and the $1$-photon state only. At resonance, both states are 
equally populated and oscillate with opposite phase. This overall yields zero
response exactly at resonance. Slightly away from resonance, one of the 
two states is slightly more populated
and a finite response arises. Further away from the resonance, the response
again
approaches zero as discussed above (see below for a quantitative evaluation). 
Hence, overall, the lineshape of an 
antiresonance arises. The $1$-photon transitions are special 
situations since only two quasienergy states are 
involved and no contributions of other quasisenergy states with smaller photon
numbers exist.  
Increasing the modulation strength, further resonances and antiresonances
appear, see Fig.\ \ref{fig.4} 
b) and c). The antiresonances turn into  resonances when the driving is further
increased. 

\begin{figure}[t]
\begin{center}
\includegraphics[width=80mm,keepaspectratio=true]{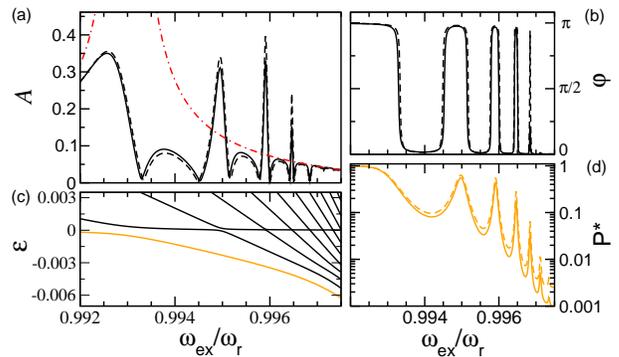} 
\caption{\label{fig.5} Solid lines: Nonlinear response of resonator in the
circuit QED 
setup based on the transmon architecture as realized in Ref.\
\onlinecite{Bishop}, i.e., 
$\omega/(2\pi) = 6.920 $ GHz, $g/\pi = 94.4 $ MHz, $\kappa/(2\pi) = 300 $ kHz,
$f/(2\pi)=8.304$ MHz, $\gamma_1/(2\pi)=98$ kHz and $\gamma_\phi=0$. Dashed
lines: same parameters, 
but without direct qubit relaxation ($\gamma_1=0$). Shown are (a) the amplitude
$A$, (b) the phase $\phi$, 
 (c) the quasienergy spectrum, and (d) the population $P^*$ of the lowest
quasienergy state $|\psi_0^*\rangle$. The red dot-dashed line in (a) shows the 
result for the population of the $0$-photon state being assumed as close to one,
see text. }
\end{center}
\end{figure}

In order to illustrate the underlying mechanism,
 we show  another example of the modulus $A$ (Fig.\ \ref{fig.5}a) and 
the phase $\varphi$ (Fig.\ \ref{fig.5}b) of the oscillator response, 
together with the associated quasienergy spectrum (Fig.\ \ref{fig.5}c) and the
occupation probability $P^*$ of the state $|\psi_0^*\rangle$ with lowest
quasienergy (Fig.\ \ref{fig.5}c). 
For the solid lines in Fig.\ \ref{fig.5}, we have chosen the same
parameters as in the fits of Ref.\ \onlinecite{Bishop}. As it turns out, in
this experiment, the qubit pure dephasing is negligible ($\gamma_\phi=0$). We
have also plotted (dashed lines) the case of
pure Purcell dissipation (no intrinsic qubit relaxation $\gamma_\phi=0$,
$\gamma_1=0$, only finite resonator damping).

First of all, we confirm that the resonances and antiresonances occur in
correspondence with the avoided crossings in the spectrum. 
As already discussed in the previous section, in absence
of resonant tunneling, i.e., 
away from any resonance, the occupation probability of the $0$-photon state is
close to one (under 
the conditions discussed above), i.e., $\rho_{00}\simeq 1$. Then, from Eq.\ 
(\ref{a00}) follows that $\langle a \rangle \approx a_{00}=\langle
0(f)|a|0(f)\rangle$, which is shown 
as red dashed line in Fig.\ \ref{fig.5}a). 
 The first antiresonance occurs for $\omega_{\rm ex}=\omega_r-g$ and has
already been discussed above. Figs.\  \ref{fig.5}a)  also shows the
$2-,3-,4-$photon resonances and the $5-,6-$photon 
antiresonances. The overall phase of the stationary oscillations changes in
presence of a resonance (Figs.\  \ref{fig.5}b), which points to a significant 
population of a state oscillating out of phase. In fact, the occupation
probability $P^*$ of the state $|\psi_0^*\rangle$ with lowest
quasienergy displays peaks at such resonant frequencies, see Fig.\ 
\ref{fig.5}d).

This phenomenology is consistent with the picture of the dissipative dynamics
drawn in the previous section.
At a multiphoton resonance, in presence of coherent resonant dynamical  
tunneling or when the rate of incoherent resonant dynamical tunneling is of
the order of the rate for an escape from the quasienergy well, a sizeable  
occupation probability of the states confined in the quasipotential well
results. 
Those quasienergy states oscillate  out of phase and with a large amplitude,
yielding 
peaks in the nonlinear response and an overall out-of-phase response
characteristics.  
 In the opposite limit of incoherent resonant dynamical tunneling 
and when no deep quasienergy well exists (i.e., in the perturbative regime) 
or when the escape rate is smaller 
than the incoherent tunneling rate, the $0$-photon state is again dominantly
populated and a small but finite occupation probability of at least one state
oscillating out of phase leads to a reduction of the resonant nonlinear response
and thus to a dip in the lineshape, but does not necessarily change the phase.  
We note that the underlying mechanism is exactly the same as for the
multiphoton transitions in the quantum Duffing oscillator
\cite{Peano1,Peano2,Peano3}.

\section{Purcell-limited set ups}

The comparison of the solid and dashed lines in Figs.\ 
\ref{fig.5}a, b, d) shows that a weak intrinsic coupling of the qubit to the
environment does not change the 
underlying physics qualitatively. In particular, transmon qubits are
Purcell-limited in the resonant regime \cite{Houck}. 
It is therefore interesting to consider the special case of pure Purcell
dissipation
$\gamma_1=\gamma_\phi=0$. In this limit, the dissipative dynamics is governed
by 
 the simplified Liouville operator
\begin{eqnarray}
{\cal L}[\rho]&\equiv&-i[H,\rho]+\frac{\kappa}{2}([a\rho,a^\dagger]+[a,\rho
a^\dagger]).
\label{PurcellME}
\end{eqnarray}
We  distinguish between two different regimes: i) when only few photons are
exchanged and no state is localized in the quasienergy well, a perturbative
analysis (in the driving)
 is in order. ii) In the opposite case, our physical picture in terms of
quasienergy surfaces is necessary to discuss the dissipative dynamics. 

\subsection{Small photon number: perturbative regime}
 
When no state is localized in the quasienergy well and away from a any
multiphoton resonance, the
quasienergy levels match the unperturbed result of Eq.\ (\ref{dsquasienergies})
and the corresponding states can be identified with the dressed states
 $|\phi_{n-}\rangle$. 

Close to the $N-$photon resonance, two scenarios are possible, as discussed
above: (A) coherent resonant dynamical tunneling (i.e., when the 
lifetime $\Gamma_N$ of the $N-$photon state $|\phi_{N \pm}\rangle$ is much
smaller than the Rabi frequency $\Omega_N$), and, (B) incoherent resonant
dynamical tunneling (i.e., when $\Gamma_N \gg \Omega_N$). 

When the dynamical tunneling is coherent ($\Gamma_N \ll \Omega_N$), the system 
tunnels many times between the states $|\phi_0\rangle$ and $|\phi_{N-}\rangle$
with period $2\pi/\Omega_N$ before it substantially decays to
$|\phi_{N-1-}\rangle$. 
The rate of this decay can be obtained in secular 
approximation as $\Gamma_N/2=\kappa(N-1/2)/2$. Subsequent decays 
from  $|\phi_{n -}\rangle$ to $|\phi_{n-1-}\rangle$ occur along the 
ladder $N-1\to N-2 \to ... \to 1 \to 0$
with decay rates ${\cal D}_{n-,(n-1)-}=(\sqrt{n}+\sqrt{n-1})^2\kappa/4$ for
$n\neq 1$ and ${\cal D}_{1-,0}=\kappa/2$. Hence, the rate from
 the $1$-photon to the $0$-photon state is smallest. 
 Note that the probability of a decay to a state with opposite dressed spin is 
small, i.e., 
${\cal D}_{(n-1)-, n+}/{\cal D}_{(n-1)+, n+}\simeq 
1/[16 n \left(n-1/2\right)]$. Thus, in this scenario, the occupation probability
of 
 $\rho_{11}$ is the largest. It can be computed (once the spin-flips are
neglected) by straightforwardly solving the master equation, which yields
\begin{equation}\label{r11}
\rho_{11}=\left(1+f^{-2}_+(N)+\sum_{n=2}^{N-1}\frac{f^{-2}_+(n)}{2}\right)^{-1}
\, ,
\end{equation}
where $f_+$ is defined below Eq.\ (\ref{fplusminus}). 
Therefore, the contribution $\rho_{11}a_{11}$ with 
\begin{equation} a_{11} = \frac{f}{2(\varepsilon_{1-}-\varepsilon_{2-})}
\frac{3+2\sqrt{2}}{4} + 
\frac{f}{4(\varepsilon_{1-}-\varepsilon_{0})} 
\label{a11}
\end{equation} 
is the largest in Eq.\ (\ref{overalloscillation}). For all multiphoton
transitions with photon number $N<6$, 
$\varepsilon_{1-}<\varepsilon_{2-},\varepsilon_0$, resulting in
the contribution $\rho_{11}a_{11}<0$ and thus in an
overall  stationary oscillation which is out of phase with the modulation.

\begin{figure}
\begin{center}
\includegraphics[width=85mm,keepaspectratio=true]{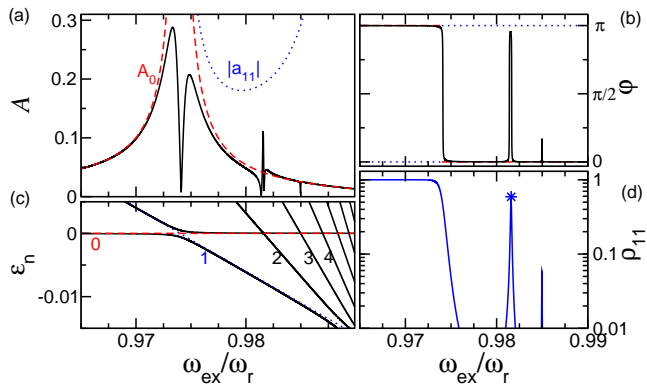}
\caption{\label{fig.6}Nonlinear response of the driven JC model: (a) amplitude,
(b) phase,  (c) quasienergies, and (d) population of the lowest quasienergy
state for $f=0.057 g ,
\kappa= 0.0009 g$ and $g=0.026 \omega_{\rm r}$. Dashed red (dotted blue) line in
(a,b,c): lowest-order
result for $|0(f)\rangle$ ($|1(f)\rangle$).}
\end{center}
\end{figure}

In the example considered in Fig.\ \ref{fig.6}  ($f=0.057 g$,  $\kappa= 0.0009
g$ and  
$g=0.007\omega_r$), this scenario is realized close to the $2-$photon
transition.
 The ratio between the $2-$photon Rabi frequency $\Omega_2=\sqrt{2}f^2/g$ (see
Eq.\ (\ref{Rabifreq})) and the lifetime of the $2-$photon dressed state
$\Gamma_2 =3/2\kappa$
is $\Omega_2/\Gamma_2 =(2\sqrt{2}/3)f^2/(g\kappa)=3.4$ and the tunneling is
coherent. 
As expected, the overall behavior changes from an in-phase oscillation with
amplitude $|a_{00}|$  (red dashed line in Fig.\ \ref{fig.6}) to a larger
out-of-phase oscillation of the order of 
$|a_{11}|$ (blue dotted line). Moreover, the occupation probability $\rho_{11}$
displays a peak (Fig. \ref{fig.6}d), whose height is close to the approximate
value computed in Eq.\ (\ref{r11}). It is represented as a blue star in Fig.
\ref{fig.6}d).

In the opposite limit  of incoherent resonant dynamical tunneling
($\Gamma_N \gg\Omega_N$), 
the population of $|\phi_0\rangle$ is $\simeq 1$,
because relaxation to the $0-$photon state is more efficient 
than any escape from there via tunneling. Dissipation then completely
washes out the resonance, and the response is identical to that away from 
 resonance and thus is in phase with the drive.  

In the intermediate regime when $\Gamma_N \simeq \Omega_N$, 
a small population $\rho_{11}$ emerges,
contributing $\rho_{11}a_{11}$ with  
$ a_{11}
<0$ given in Eq.\ (\ref{a11}), 
which is negative for $N<6$ (in the perturbative regime). This also 
leads to a reduced response with an antiresonance. This scenario is realized
close to the $3-$photon transition in Fig. \ref{fig.6}). In this case, the Rabi
frequency is
$\Omega_3=(3^{7/2}/2^4) f^3/g^2$ and $\Gamma_3=(5/2)\kappa$, so that
$\Omega_3/\Gamma_3=0.24$.

\subsection{Large photon number: semiclassical regime}
\label{sec.semicl}
When several photons are contained in the resonator, a picture in terms of 
 the dissipative semiclassical dynamics emerges. As will be shown
below, a separation of relaxation time scales exists, 
 which separates fast intrawell from 
slow interwell relaxation.
We first focus on the intrawell dynamics. Close to the minimum, the wave
functions are given by Eq.\ (\ref{lowestenergystate2}). We can compute the
corresponding
dissipative transition rates by plugging them into Eq.\ (\ref{PurcellME}).
In this limit, dissipative transitions occur only between nearest neighbors,
with the rates
${\cal D}_{n-1,n}=\kappa n \cosh^2 r$ and
 ${\cal D}_{n,n-1}=\kappa n\sinh^2r$.
Here,  the detailed balance condition is fulfilled. 
Hence, when the system is initially in a state with a 
large photon number $n$, it falls with 
large probability into the basin of attraction of the 
 quasienergy minimum. This process constitutes intrawell relaxation. 
When furthermore damping is smaller than the intrawell 
level spacing, i.e., $\kappa\ll g\lambda\omega^*$ (see Eq.\ (\ref{harmen})), 
detailed balance is retained and determines an 
effective Boltzmann distribution
\begin{equation}\label{boltzman}
 P^*_n=P^* e^{-n \beta_{\rm eff}}\,,
\end{equation}  with the 
 effective inverse temperature $\beta_{\rm eff}=2 \ln \coth r^*$  being defined
in terms of 
the squeezing parameter $r^*$.  We emphasize that this link between effective
temperature and squeezing can be
generalized to any driven quantum system with a smooth 
 quasienergy surface and coupled linearly to a bath (e.g., the linearly and the
parametrically driven Duffing
oscillator). 
It can be easily generalized to finite real temperatures $T>0$ as well. It turns
out
that the zero temperature limit applies  when 
$\sinh^2 r$  is much larger than the bosonic occupation number 
$\bar{n}(\omega_{\rm ex}/T)$ of the bath taken at frequency $\omega_{\rm ex}$. 
In the opposite limit, $\beta_{\rm eff}=\omega_{\rm ex}/T$. Since we include
here only photon leaking, 
i.e., $\omega_{\rm ex} \gg T$, the effective temperature is still small.   

On the large time scale, the system can escape from the basin of attraction of
the quasienergy minimum with a small forward rate
$k_+$.   Overall there are two mechanism contributing to the escape: i) The
system can climb up the quasienergy well by quantum activation.
As already discussed in Section \ref{sec.dissdyn}, the corresponding  rate
$\Gamma_a$ is suppressed exponentially, following 
$\Gamma_a\propto \kappa e^{-c/\lambda}$. Moreover, ii)
the bath induces spin-flip transitions with rate $\Gamma_s$ even  in absence of
an intrinsic spin-bath coupling. This is a consequence of the dressing of the
spin. 
From Eqs.\ (\ref{arotated}) and (\ref{PurcellME}), it follows that 
$\Gamma_s\propto\lambda$. Therefore, this mechanism is dominant for small
$\lambda$ and the rate $k_+$ vanishes according to $k_+\propto\lambda$. In real
set-ups,  a lower bound to
$k_+$ is given by the small intrinsic couplings $\gamma_1$ and 
$\gamma_\phi$ of the spin to the bath.

\begin{figure}
\begin{center}
\includegraphics[width=85mm,keepaspectratio=true]{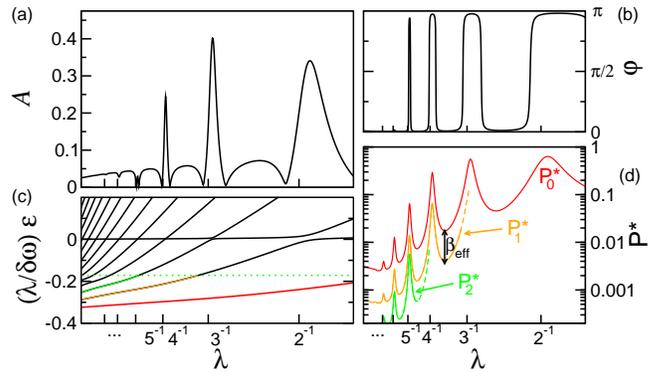}
\caption{\label{fig.7}Nonlinear response of the driven JC model: (a) amplitude,
(b) phase,  (c) quasienergies, and (d) population of the lowest quasienergy
states $|\psi^*_n\rangle$ ($n=0, 1, 2$) as a function of the rescaled 
Planck constant $\lambda$ for the parameters $f/g=0.171,
\kappa/g=6.1 10^{-3} \omega_{\rm r}$, $\gamma_1, \gamma_\phi=0$ (Purcell limit).
  Green dotted line in (c): quasienergy of the saddle point.}
\end{center}
\end{figure}

Once the system has left the basin of attraction of the bright state
$|\psi_0^*\rangle$, it quickly decays to the dark state $|0(f)\rangle$. 
From there, it can escape with a backward rate 
 $k_-$ on a large time scale. 
The stationary populations of the intrawell states, which oscillate out of 
phase, and the 
$0-$photon state, which oscillates in phase with the modulation, 
 are determined by the ratio $k_-/k_+$. 
Away from any resonance, photon leaking favors the $0-$photon (dark) state and 
thus, the corresponding interwell relaxation is fast, i.e., $k_+\gg k_-$. 
Close to a resonance for $\lambda=1/N$ (N integer), if the driving is resonant
or the incoherent resonant tunneling rate $k_-=\Omega_N^2/\Gamma_N$
 is of the same order of $k_+$, the
response 
is qualitatively modified and the resonant-antiresonant transition is now 
governed by the ratio $k_- /k_+$. 

Next, we complete our discussion with numerical results for a concrete
example. In Fig.\ \ref{fig.7}, we plot (a) the overall oscillation amplitude,
(b) the overall phase, (c) the rescaled 
quasienergy spectrum, and (d) the occupation probabilities of the first three
states $|\psi^*_n\rangle$ ($n=0, 1, 2$) at the well bottom
as a function of the effective Planck constant $\lambda$. For
$\lambda\approx1/N$ ($N=2, 3, 4, 5$), we observe a resonant out-of-phase
response (see Figs.\ \ref{fig.7} a) and b),
whereas for $\lambda\approx1/6$,   an antiresonance appears which is in phase
with the modulation signal. For the same values of $\lambda$,
 the occupation probabilities of the states in the well display peaks, see 
Fig.\ \ref{fig.7} d).  The
heights of the respective peaks are suppressed exponentially with $N$. 
According to the
detailed
balance condition in Eq. (\ref{boltzman}), the ratio of the probabilities for
nearby states close to the bottom of the well is $\exp(-\beta_{\rm eff})$.
In fact, in logarithmic scale, the individual curves of the probabilities tends
to be equidistant. 
The theoretical value for the gap between the curves is shown by the black
double arrow, indicating that the agreement with Eq.\ 
(\ref{boltzman}) is also quantitative.

\begin{figure}
\begin{center}
\includegraphics[width=65mm,keepaspectratio=true]{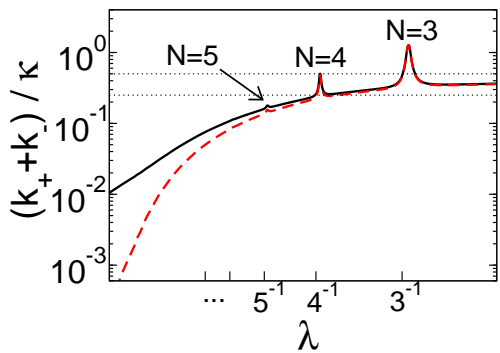}
\caption{\label{fig.8} Solid line: Smallest eigenvalue of the Lindblad master
equation. Dashed red line: 
the same without dissipative spin flips. Dashed-dotted orange line: decay rate
for $|\psi^*\rangle \to 
| \phi_{0}^f \rangle$. Dotted lines correspond to $\kappa/2$ and $\kappa/4$.} 
\end{center}
\end{figure}

In order to underpin the drawn picture by more quantitative results, it
is is helpful to consider the eigenvalues of the Liouville 
operator. For this, we start deep in the semiclassical regime, i.e., 
with small detuning. Here, 
a clear separation of time scales for the dissipative dynamics on the bistable
quasienergy surface 
occurs. Well-defined energy wells with a large quasienergy barrier in between
exist for small detuning and 
allow for a clear description in terms of a single relaxation rate (see also
Ref.\ \onlinecite{rate} for a comprehensive review). 
In this regime, we find a single eigenvalue $\Gamma$ which 
consists of the sum of $k_-$ and $k_+$ ,
is real and much smaller than the real parts of 
all the other eigenvalues. This is shown in Fig.\ \ref{fig.8} as black solid
line. In order to emphasize the role of bath induced spin-flips, we have 
also computed the smallest eigenvalue $\Gamma$ of the Liouvillian after  
removing those transition by hand from the master equation, see red dashed
line in Fig.\ \ref{fig.8}. For increasing $\lambda$, i.e., 
when $\lambda \gtrsim 0.25$,  we enter a regime, 
where the separation of time scales 
is not so clearly expressed and $\Gamma$ becomes comparable to the real parts of
the next three eigenvalues. The latter correspond to relaxation (one real
eigenvalue) and to decoherence 
(a complex conjugated pair of eigenvalues), involving the pair of states
$|\phi_{1+}\rangle$ and  $|\phi_0\rangle$. We do not 
show them in the figure, but instead show  the perturbatively determined values 
$\kappa/2$ (relaxation) and $\kappa/4$ (decoherence) out of resonance as dotted
horizontal lines. The peaks in the total rate $\Gamma$ are due to resonant
dynamical tunneling for $k_+$ as discussed in Section \ref{sectunneling}.
Due to the logarithmic  scale, peaks in $\Gamma$ are only visible when
$k_+$ is of the order or larger than the background value given by  $k_-$.

\section{Connection to other models} The analysis presented here can in general
be 
extended to any driven nonlinear oscillator coupled
bilinearly to a thermal bath. 
 For example, the Duffing oscillator is characterized by two classical stable
solutions with opposite oscillation phase. 
In the quantum regime, two quantum squeezed states correspond to them.  
For weak driving, the small-oscillation solution can be 
identified with the $0-$photon quantum state. 
At low thermal energies, it is dominantly populated away from any resonance 
due to photons leaking into the bath. Thus, it can be regarded as 
stable in absence of any multiphoton transitions. However, 
 this stable quasienergy state is associated to a relative 
quasienergy maximum, which is in direct contrast
to the case of a static bistable potential. There, the lowest energy 
state is always the minimum of the true potential.   
At a multiphoton resonance, the zero-photon quasienergy state is no 
longer stable  since excitations to a $N-$photon state occur. Hence, 
it becomes metastable and generates a (anti-)resonance of the stationary
oscillation. This behavior of the quantum Duffing oscillator has been already
predicted \cite{Peano1,Peano2,Peano3}, but the link to the
semiclassical picture \cite{Dmitriev86,Dykman88}
has not been drawn. 
Thus, the generic behavior of a driven damped nonlinear quantum 
oscillator includes dynamically generated metastable states from which 
the system can escape via thermal diffusion, quantum activation or dynamical 
tunneling. In the regime of many photons in the resonator, the escape rate of 
dynamical tunneling processes can be obtained in a semiclassical description,
while 
in the regime of only few photons (deep quantum regime), the escape can occur
via 
resonant dynamical tunneling, leading to resonant multiphoton transitions.
Depending 
on the ratio of the Rabi frequency and the lifetime of the multiphoton state,
the resonant 
dynamical tunneling can be coherent (for large quality factors) or incoherent
(for small quality factors).  Depending on the phase of the associated
multiphoton transitions, a resonant or an antiresonant nonlinear response may
arise.  Such a situation is also expected in a Josephson bifurcation amplifier
\cite{jba} operated in the deep quantum regime (note that all related
experimental setups realized so far operate in the classical regime).

\section{Conclusions}
In conclusion, inspired by recent experiments, we 
have shown that a driven circuit QED setup can acquire 
 a dynamical bistability. The relevant model to describe this 
is the driven dissipative Jaynes-Cummings model for which we have analyzed 
its nonlinear response properties. We have shown that a quasienergy surface can
be 
derived in a rotating frame picture which clearly shows two metastable basins
of 
attraction. This picture is also convenient for studying the semiclassical 
limit. We have predicted the existence of a
metastable quantum squeezed state in the semiclassical limit and have discussed 
 a connection between 
effective local temperature at the bottom of the quasienergy well 
and the squeezing parameter. We have analyzed the
escape mechanisms from the metastable states and 
found resonant dynamical tunneling, both in an incoherent and a coherent
version. 
Our analysis adds another example to the series of 
nonlinear driven dissipative quantum systems with surprising and
counterintuitive, but 
generic features. The thorough experimental investigation of these intrinsic
quantum effects  
is an interesting prospect.

\acknowledgments

We thank M.\ I.\ Dykman and V.\ Leyton  for useful discussions
and acknowledge support by 
 the Excellence Initiative of the German Federal and State Governments.


\begin{thebibliography}{99}
\bibitem{Nayfeh}A.H.\ Nayfeh and D.T. Mook, {\it Nonlinear Oscillations} 
(Wiley, New York, 1979).  

\bibitem{Dykman80}M.I. Dykman and M.A. Krivoglaz, Physica A
{\bf 104}, 480 (1980).

\bibitem{Dykman79}M.I. Dykman and M.A. Krivoglaz, Sov. Phys. JETP {\bf 50}, 
30 (1979).  

\bibitem{Dykman90}M.I. Dykman and V.N. Smelyanski, Phys. Rev. A {\bf 41}, 
3090 (1990).

\bibitem{Dmitriev86}A.P. Dmitriev and M.I. D'yakonov, Sov. Phys. JETP {\bf 63},
838 (1986).

\bibitem{Dykman88}M.I. Dykman and V.N. Smelyanskii, Sov. Phys. JETP {\bf 67},
1769 (1988). 

\bibitem{Dykman06}M. Marthaler and M.I. Dykman, Phys.\ Rev.\ A  {\bf 73}, 042108
(2006). 

\bibitem{Dykman07}C. Hicke and M.I. Dykman, Phys. Rev. B {\bf 76}, 054436
(2007). 

\bibitem{Peano1}V. Peano and M. Thorwart, Phys. Rev. B {\bf 70}, 235401 (2004).

\bibitem{Peano2}V. Peano and M. Thorwart, Chem. Phys. {\bf 322}, 135 (2006).

\bibitem{Peano3}V. Peano and M. Thorwart, New J. Phys. {\bf 8}, 21 (2006). 

\bibitem{Peano10}V. Peano and M. Thorwart, Europhys. Lett. {\bf 89}, 17008
(2010). 

\bibitem{jaynescummings} E.T. Jaynes and F.W. Cummings, Proc. IEEE {\bf 51}, 89
(1963)

\bibitem{Larson}J. Larson, Phys. Scr. {\bf 76}, 146 (2007).

\bibitem{Wallraff}A. Wallraff {\em et al.}, 
Nature {\bf 431}, 162 (2004).

\bibitem{Chiorescu}I. Chiorescu {\em et al.}, 
Nature {\bf 431}, 159 (2004).

\bibitem{Hatakenaka}N. Hatakenaka and S. Kurihara, Phys. Rev. A {\bf 54}, 1729
(1996). 

\bibitem{Winger}M. Winger {\em et al.}, 
Phys.\ Rev.\ Lett.\ {\bf 101}, 226808 (2008). 

\bibitem{Blais04}A. Blais, R.-S. Huang, A. Wallraff, S.M. Girvin, and R.J.
Schoelkopf, Phys. Rev. A {\bf 69}, 062320 (2004). 

\bibitem{Moon05}K. Moon and S.M. Girvin, 
Phys. Rev. Lett. {\bf 95}, 140504 (2005).

\bibitem{Wallquist06} M. Wallquist, V.S. Shumeiko, and G. Wendin, 
 Phys. Rev. B {\bf 74}, 224506 (2006). 

\bibitem{Ficek02}S. Swain and Z. Ficek, J. Opt. B: Quantum Semiclass. Opt. 
{\bf 4}, S328 (2002). 

\bibitem{Hauss08}J. Hauss, A. Fedorov, C. Hutter, A. Shnirman, and G. Sch\"on, 
Phys. Rev. Lett. {\bf 100}, 037003 (2008).

\bibitem{Xiao}R.J. Brecha, P.R. Rice, and M. Xiao, Phys. Rev. A {\bf 59}, 2392 
(1999). 

\bibitem{Bishop} L.S. Bishop {\em et al.}, 
Nature Phys.\ {\bf 5},  105 (2009).

\bibitem{Carmichael}L. Tian and H.J. Carmichael, 
Phys.\ Rev.\ A {\bf 46}, 6801 (R) (1992). 

\bibitem{Schoelkopf} R.J. Schoelkopf and S. M. Girvin, 
Nature {\bf 451}, 664 (2008). 

\bibitem{Thomson} R. J. Thomson, G. Rempe, and H. J. Kimble, Phys.\ Rev.\ Lett.
{\bf 68}, 1132  (1992). 

\bibitem{Haroche} S. Haroche and J.-M. Raimon, 
  {\it Exploring the Quantum},
   (Oxford University Press, Oxford 2006).

\bibitem{note} The terms neglected in the RWA could be taken care in the
framework of Floquet theory. They are of the order of $g/\omega_{\rm ex}$ and
$f/\omega_{\rm ex}$.

\bibitem{persico} P. Carbonaro, C.\ Compagno and F.\ Persico, Phys. Lett. {\bf
73 a}, 99 (1979) 

\bibitem{note2}  The contributions to the quasienergies of order $\lambda$ from
higher order terms of the quasipotential should be taken into account but cancel
out. 

\bibitem{yuen} H. P. Yuen, Phys. Rev. A {\bf 13}, 2226 (1976).

\bibitem{Scholkopf} M. D. Reed {\em et al.}, arXiv:1004.4323

\bibitem{Girvin} L. Bishop, E. Ginossar  and S. M. Girvin, arXiv:1005.0377

\bibitem{Milburn} G. Milburn and P. Alsing, in 
  {\it Directions in Quantum Optics, Lecture Notes in Physics},
  ed. by H.J. Carmichael, R.J. Glauber, and M.O. Scully, 
  {\bf 561}, 303 (Springer, Berlin 2001).

\bibitem{Unruh} W. Unruh, Phys. Rev. D {\bf 14}, 870 (1976).

\bibitem{Houck} A. A. Houck  {\em et al.}, 
Phys.\ Rev.\ Lett.\ {\bf 101}, 080502 (2008). 

\bibitem{rate} P.\ H\"anggi, P.\ Talkner, and M.\ Borkovec,
  Rev. Mod. Phys. {\bf 62}, 251 (1990).

\bibitem{jba} I. Siddiqi {\em et al.}, 
Phys.\ Rev.\ Lett.\ {\bf 93}, 207002 (2004).



\end{thebibliography}
\end{document}